\documentclass[aps,floats,showpacs,preprint]{revtex4} 
\usepackage{feynmp}
\usepackage{graphicx}
\usepackage{epsfig}
\usepackage{psfig}
\usepackage{amsmath}
\usepackage{hyperref}

\def\PRD#1{{ Phys.\ Rev.} {\bf D#1}}
\def\NPB#1{{ Nucl.\ Phys.} {\bf B#1}}
\def\PLB#1{{Phys.\ Lett.} {\bf B#1}}

\def\be{\begin{equation}}
\def\ee{\end{equation}}
\def\bea{\begin{eqnarray}}
\def\eea{\end{eqnarray}}

{\newcommand{\lsim}{\mbox{\raisebox{-.6ex}{~$\stackrel{<}{\sim}$~}}}
{
\def\sss{\scriptscriptstyle}

\begin{document}

\title{\bf Leptogenesis bound on neutrino masses in left-right 
symmetric models with spontaneous D-parity violation} 

\author{Narendra Sahu}
\email{narendra@prl.res.in}
\author{Utpal Sarkar}
\email{utpal@prl.res.in}
\affiliation{Theoretical Physics Division, Physical Research 
Laboratory, Ahmedabad-380 009, India}

\begin{abstract}
We study the baryogenesis via leptogenesis in a class of left-right 
symmetric models, in which $D$-parity is broken spontaneously.
We first discuss the consequence of the spontaneous 
breaking of $D$-parity on the neutrino masses. Than we
study the lepton asymmetry in various cases, from the decay of right 
handed neutrino as well as the triplet Higgs, depending on their relative 
masses they acquire from the symmetry breaking pattern. The leptogenesis 
bound on their masses are discussed by taking into account the low energy 
neutrino oscillation data. It is shown that a TeV scale leptogenesis is 
viable if there is an additional source of $CP$ violation like 
$CP$-violating condensate in the left-right domain wall. This is 
demonstrated in a class of left-right symmetric models where 
$D$-parity breaks spontaneously at a high energy scale while allowing 
$SU(2)_R$ gauge symmetry to break at the TeV scale.
\end{abstract}
\maketitle

\section{Introduction}
\noindent
The matter antimatter asymmetry during the big-bang nucleosynthesis 
era is required to be very tiny. Recent results from the 
Wilkinson Microwave Anisotropy Probe (WMAP) provides a fairly
precise value for this asymmetry, given by~\cite{spergel.03} 
\be
\left( \frac{(n_B-n_{\bar{B}})}{n_\gamma} \right)_0 \equiv 
\left( \frac{n_B}{n_\gamma} \right)_0=\left(6.1^{+0.3}_{-0.2}\right)
\times 10^{-10}.
\label{asymmetry}
\ee

In recent years the most fascinating experimental result in
particle physics came out in neutrino physics. The atmospheric
neutrinos provided us the first evidence for a non-vanishing neutrino
mass~\cite{atmos-data} and hence first indication for physics
beyond the standard model. The mass-squared difference 
providing $\nu_\mu-\nu_\tau$ oscillations, as required by the 
atmospheric neutrinos, is given by
\be
\Delta m_{atm} \equiv \sqrt{|m_3^2-m_2^2|}\simeq 0.05 eV\,.
\label{atmos-scale} 
\ee
This result is further strengthened by the solar neutrino 
results~\cite{solar-data} which require a mass-squared 
difference providing a $\nu_e-\nu_\mu$ oscillation. The mass 
splitting given by
\be
\Delta m_{\odot}\equiv \sqrt{m_2^2-m_1^2}\simeq 0.009 eV\,, 
\label{solar-scale}
\ee
where $m_1$, $m_2$ and $m_3$ are the masses of light physical 
neutrinos. Note that $\Delta m_\odot$ is positive as indicated by 
the SNO data while there is an ambiguity in the sign of 
$\Delta m_{atm}$ to the date.  

The above discoveries, the matter antimatter asymmetry of the 
present Universe (\ref{asymmetry}) and the sub-eV neutrino masses 
(\ref{atmos-scale}) and (\ref{solar-scale}), could be intricately 
related with each other. A most viable scenario to explain is the 
baryogenesis via leptogenesis (BVL)~\cite{fukugita.86,baryo_lepto_group}. 
The smallness of the neutrino masses compared to the charged fermions
are best understood in terms of a seesaw mechanism~\cite{seesaw_group}.
Although the neutrinos are massless in the standard model, a minimal
extension including right-handed neutrinos or triplet Higgs scalars
or both can generate tiny Majorana masses for the neutrinos through
the seesaw mechanism. The smallness of the neutrino masses depend
on a large suppression by the lepton (L) number violating scales in the
model, which is the scale of Majorana masses of the right-handed 
neutrinos or the masses and dimensional couplings of the triplet Higgs 
scalars. The $L$-number violating decays of the right-handed neutrinos 
or the triplet Higgs scalars at this large scale can then generate a
$L$-asymmetry of the universe, provided there is enough
$CP$-violation and the decays satisfy the out-of-equilibrium condition,
the necessary criteria of Sakharov~\cite{sakharov.67}. This $L$-asymmetry 
of the universe is then get converted to a baryon (B)
asymmetry of the universe (BAU) through the sphaleron processes 
unsuppressed above the electroweak phase transition~\cite{krs_plb.85}. 

In the simplest type-I seesaw models the singlet right-handed 
neutrinos ($N_R$'s) are added to the Standard Model ($SM$) gauge 
group, $SU(2)_L\times U(1)_Y$. The canonical seesaw then 
gives the light neutrino mass matrix: 
\be
m_\nu=m_\nu^I=-m_DM_R^{-1}m_D^T\,,
\label{typeI-seesaw}
\ee
where $m_D$ is the Dirac mass matrix of the neutrinos connecting 
the left-handed neutrinos with the right-handed neutrinos and
$M_R$ is the Majorana mass matrix of the right handed heavy neutrinos,
which also sets the scale of $L$-number violation. Since the Majorana 
mass of the right handed neutrinos violate $L$-number by two units, their 
out of thermal equilibrium decay to $SM$ particles is a natural source of 
$L$-asymmetry~\cite{fukugita.86}. The $CP$-violation, which comes from the 
Yukawa couplings that gives the Dirac mass matrix, resulted through the 
one loop radiative correction requires at least two right handed neutrinos. 
Assuming a strong hierarchy in the right handed neutrino sector a successful 
$L$-asymmetry in these models requires the mass scale of the lightest right 
handed neutrino to be $M_1\geq O(10^{9})$ GeV~\cite{type-I-bound}. If 
the corresponding theory of matter is supersymmetric then this bound, 
dangerously being close to the reheat temperature, poses a problem. A 
modest solution was proposed in ref.~\cite{ma&sahu.06} by introducing an 
extra singlet. However, the success of the model is the reduction of above 
bound~\cite{type-I-bound} by an order of magnitude. 

In the type-II seesaw models, on the other hand, triplet Higgses
($\Delta_L$'s) are added to the $SM$ gauge group. The triplet
seesaw~\cite{tripletseesaw_group} in this case gives the light 
neutrino mass matrix:
\be
m_\nu=m_\nu^{II}=f\mu \frac{v^2}{M_{\Delta_L}^2}\,,
\label{typeII-seesaw}
\ee
where $M_{\Delta_L}$ is the mass of the triplet Higgs scalar 
$\Delta_L$, $f$ is the Yukawa coupling relating the triplet Higgs
with the light leptons, $\mu$ is the coupling constant with mass 
dimension 1 for the trilinear term with the triplet Higgs and two standard
model Higgs doublets and $v$ is the vacuum expectation value ($vev$) of 
the $SM$ Higgs doublet. The $L$-asymmetry, in these models, is generated 
through the $L$-number violating decays of the $\Delta_L$ to $SM$ lepton 
and Higgs. The $CP$-violation, originated from the one loop radiative 
correction, requires at least two triplets. Again the scale of $L$-number 
violation is determined by $M_{\Delta_L}$ and $\mu$ and required to be very 
high and larger than the type-I models~\cite{type-II-group}. 

An attractive scenario is the hybrid seesaw models (type-I+type-II), 
where both right-handed neutrino as well as triplet Higgs scalar are 
present. So, there is no constraint on their number to have 
$CP$-violation. The neutrino mass matrix in these models is given by 
\be
m_\nu=m_\nu^I+m_\nu^{II}\,,
\label{hybrid-seesaw}
\ee
where $m_\nu^I$ and $m_\nu^{II}$ are given by equations (\ref{typeI-seesaw}) 
and (\ref{typeII-seesaw}) respectively. A natural extension of the $SM$ to 
incorporate both type-I as well as type-II terms of the neutrino mass 
matrix is the left-right symmetric models~\cite{left-right-group} with
the gauge group $SU(2)_L\otimes SU(2)_R\otimes U(1)_{B-L}$.
The advantages of considering this model is that (1) it has a natural
explanation for the origin of parity violation, (2) it can be easily
embedded in the $SO(10)$ Grand Unified Theory (GUT) and (3) $B-L$ is
a gauge symmetry. Since $B-L$ is a gauge symmetry of the model, it
is not possible to have any $L$-asymmetry before the left-right
symmetry breaking. An $L$-asymmetry can be produced after the left-right
symmetry breaking phase transition, either through the decay of right handed
neutrinos or through the decay of the triplet Higgses or can be both
depending on the relative magnitudes of their masses. Assuming a strong 
hierarchy in the right-handed neutrino sector and $M_1<M_{\Delta_L}$, it 
is found that $M_1$ can be reduced to an order of magnitude in 
comparison to the type-I models~\cite{hamby&sen_plb, 
antusch&king_plb, type-II-bound}. Despite the success, this mechanism 
of producing $L$-asymmetry in these models can not bring down the
scale of leptogenesis to the scale of the next generation 
accelerators. 

The alternatives to these are provided by mechanisms which work at 
the TeV scale~\cite{susy_tev} either in supersymmetric extensions of 
the $SM$ relying on the new particle content or finding the additional source 
of CP violation in the model~\cite{extra_tev}. It is worth investigating 
other possibilities, whether or not supersymmetry is essential to the 
mechanism. In the following we consider a class of left-right symmetric
models in which the spontaneous breaking of $D$-parity occurs at a
high energy scale ($\sim 10^{13} GeV$) leaving the $SU(2)_R$ intact. 
In the left-right symmetric models, parity connects the left-handed
gauge group with the right-handed gauge group. But the same need not
be true for the scalar particles. In this class of left-right symmetric 
models, the spontaneous $D$-parity
violation allows the scalars transforming under the group $SU(2)_L$
to decouple from the scalars transforming under the group $SU(2)_R$
and these scalars can have different masses and couplings. 
This allows the mass scale of the triplet $\Delta_L$ to be very high
at the $D$-parity breaking scale~\cite{paridaetal.84} while leaving the mass 
of $\Delta_R$ to be as low as the $SU(2)_R$ symmetry breaking scale
or vice versa. However, we will see that even in these models
a successful leptogenesis doesn't allow neither the mass of triplets nor 
the mass of right handed neutrinos less than $10^8$ GeV if the 
$L$-asymmetry arises from their out of equilibrium decay. We then consider an 
alternative mechanism to bring down the mass scale of right handed 
neutrinos to be in TeV scale. In the respective mechanism 
a net $L$-asymmetry arises through the preferential scattering 
of left-handed neutrino $\nu_L$ over its CP conjugate state $\nu_L^c$ 
from the left-right domain wall~\cite{cline&yajnik_prd.02}. The 
survival of this asymmetry then requires the mass scale of lightest 
right handed neutrino, assuming a normal mass hierarchy in the right 
handed neutrino sector, to be in TeV scale~\cite{sahu&yaj_prd.05,
sahu&yaj_plb.06}. In this class of models the TeV scale masses of 
the right handed neutrinos are resulted through the low scale 
($\sim 10$ TeV) breaking of $SU(2)_R$ gauge symmetry while $D$-parity 
breaks at a high energy scale ($\sim 10^{13}$ GeV). This is an 
important result pointed out in this paper.

The rest of the manuscript is arranged as follows. In the section-II 
we briefly discuss the left-right symmetric models, elucidating 
the required Higgs structure for spontaneous breaking of $D$-parity. 
In section-III we discuss the parities in left-right symmetric models 
and their consequence on neutrino masses. Than we give a possible path 
for embedding the left-right symmetric models in the $SO(10)$ GUT. In 
section-IV we discuss the production of $L$-asymmetry through the 
decay of heavy Majorana neutrinos as well as the triplet $\Delta_L$ 
separately by taking into account the relative magnitudes of their 
masses. In section V, by assuming a charge-neutral symmetry, we 
derived the neutrino mass matrices from the low energy neutrino 
data. Using this symmetry the $L$-asymmetry is estimated in 
section VI by considering the relative masses of $N_1$ and the 
triplet $\Delta_L$. In any case, it is found that the leptogenesis 
scale can not be lowered to a scale that can be accessible in the 
next generation accelerators. In section VII, we therefore discuss 
an alternative mechanism which has the ability to explain the 
$L$-asymmetry at the TeV scale. In section VIII we give a qualitative 
suggestion towards the density perturbations due to the presence of 
heavy singlet scalars. We summarize our results and conclude 
in section IX.  

\section{Left-Right symmetric models}
In the Left-Right symmetric model, the right handed charged lepton 
of each family which was a isospin singlet under $SM$ gauge group 
gets a new partner $\nu_R$. These two form an isospin doublet 
under the $SU(2)_R$ of the left-right symmetric gauge group 
$SU(2)_L\times SU(2)_R\times U(1)_{B-L}\times P$, where $P$ 
stands for the parity. Similarly, in the quark-sector, the right 
handed up and down quarks of each family, which were isospin singlets 
under the $SM$ gauge group, combine to form the isospin doublet under 
$SU(2)_R$. As a result before the left-right symmetry breaking both 
left and right handed leptons and quarks enjoy equal strength of 
interactions. This explains that the parity is a good quantum number 
in the left-right symmetric model in contrast to the $SM$ where the left 
handed particles are preferential under the electro-weak interaction. 

In the Higgs sector, the model consists of a $SU(2)$ singlet scalar 
field $\sigma$, two $SU(2)$ triplets $\Delta_L$ and $\Delta_R$ and 
a bidoublet $\Phi$ which contains two copies of $SM$ Higgs. Under 
$SU(2)_L\times SU(2)_R \times U(1)_{B-L}$ the field contents and the 
quantum numbers of the Higgs fields are given as
\bea
\sigma &\sim& (1,1,0)\\
\Phi &=&\begin{pmatrix}
\phi_{1}^{0} & \phi_{1}^{+}\\
\phi_{2}^{-} & \phi_{2}^{0}
\end{pmatrix} \sim (2,2,0)
\eea
\bea
\Delta_{L} &=& \begin{pmatrix}
\delta_{L}^{+}/\sqrt{2} & \delta_{L}^{++}\\
\delta_{L}^{0} & -\delta_{L}^{+}/\sqrt{2}
\end{pmatrix}\sim (3,1,2)\\
\Delta_{R} &=& \begin{pmatrix}
\delta_{R}^{+}/\sqrt{2} & \delta_{R}^{++}\\
\delta_{R}^{0} & -\delta_{R}^{+}/\sqrt{2}
\end{pmatrix}\sim (1,3,2).
\eea

The most general renormalizable Higgs potential exhibiting left-right 
symmetry is given by~\cite{left-right-potential} 
\be
\mathbf{V}=\mathbf{V}_{\sigma}+\mathbf{V}_{\Phi}+\mathbf{V}_{\Delta}
+\mathbf{V}_{\sigma\Delta}+\mathbf{V}_{\sigma\Phi}+\mathbf{V}_{\Phi\Delta}\,,
\label{potential}
\ee
where 
\be
\mathbf{V_\sigma} = -\mu_{\sigma}^2 \sigma^2+\lambda_\sigma 
\sigma^4\nonumber\,,
\label{sigma-pot}
\ee
\begin{eqnarray*}
\mathbf{V}_\Delta &=&-\mu _{\Delta}^2\left[ Tr\left( \Delta _L\Delta _L^{\dagger
}\right) +Tr\left( \Delta _R\Delta _R^{\dagger }\right) \right] \\
&&+\rho _1\left\{ \left[ Tr\left( \Delta _L\Delta _L^{\dagger }\right)
\right] ^2+\left[ Tr\left( \Delta _R\Delta _R^{\dagger }\right) \right]
^2\right\} \\
&&+\rho _2\left[ Tr\left( \Delta _L\Delta _L\right) Tr\left( \Delta
_L^{\dagger }\Delta _L^{\dagger }\right) +Tr\left( \Delta _R\Delta _R\right)
Tr\left( \Delta _R^{\dagger }\Delta _R^{\dagger }\right) \right] \\
&&+\rho _3\left[ Tr\left( \Delta _L\Delta _L^{\dagger }\right) Tr\left(
\Delta _R\Delta _R^{\dagger }\right) \right] \\
&&+\rho _4\left[ Tr\left( \Delta _L\Delta _L\right) Tr\left( \Delta
_R^{\dagger }\Delta _R^{\dagger }\right) +Tr\left( \Delta _L^{\dagger
}\Delta _L^{\dagger }\right) Tr\left( \Delta _R\Delta _R\right) \right] ,
\label{delta-pot}
\end{eqnarray*}
\begin{eqnarray*}
\mathbf{V}_\Phi &=&-\mu _{\Phi1}^2 Tr\left( \Phi ^{\dagger }\Phi \right) -\mu
_{\Phi2}^2\left[ Tr\left( \widetilde{\Phi }\Phi ^{\dagger }\right) +Tr\left(
\widetilde{\Phi }^{\dagger }\Phi \right) \right] \\
&&+\lambda _1\left[ Tr\left( \Phi \Phi ^{\dagger }\right) \right] ^2+\lambda
_2\left\{ \left[ Tr\left( \widetilde{\Phi }\Phi ^{\dagger }\right) \right]
^2+\left[ Tr\left( \widetilde{\Phi }^{\dagger }\Phi \right) \right]
^2\right\} \\
&&+\lambda _3\left[ Tr\left( \widetilde{\Phi }\Phi ^{\dagger }\right)
Tr\left( \widetilde{\Phi }^{\dagger }\Phi \right) \right] \\
&&+\lambda _4\left\{ Tr\left( \Phi ^{\dagger }\Phi \right) \left[ Tr\left(
\widetilde{\Phi }\Phi ^{\dagger }\right) +Tr\left( \widetilde{\Phi }%
^{\dagger }\Phi \right) \right] \right\} ,
\end{eqnarray*}
\begin{eqnarray*}
\mathbf{V}_{\sigma \Delta} &=& M\sigma \left[ Tr(\Delta_L \Delta_L^\dagger) - 
Tr(\Delta_R \Delta_R^\dagger) \right]+\gamma \sigma^2 \left( 
Tr(\Delta_L \Delta_L^\dagger) + Tr(\Delta_R\Delta_R^\dagger)\right) ,
\label{sigma-delta}
\end{eqnarray*}
\begin{eqnarray*}
\mathbf{V}_{\sigma\Phi} &=& \delta_1 \sigma^2 Tr(\Phi^\dagger\Phi)+ 
M'\sigma \left[ Tr(\tilde{\Phi}\Phi^\dagger) - Tr(\tilde{\Phi}^\dagger\Phi) 
\right]\nonumber\\
&+& \delta_2 \sigma^2\left[ Tr(\tilde{\Phi}\Phi^\dagger)+ 
Tr(\tilde{\Phi}^\dagger\Phi) \right] ,
\end{eqnarray*}
\begin{eqnarray*}
\mathbf{V}_{\Phi \Delta } &=&\alpha _1\left\{ Tr\left( \Phi ^{\dagger }\Phi
\right) \left[ Tr\left( \Delta _L\Delta _L^{\dagger }\right) +Tr\left(
\Delta _R\Delta _R^{\dagger }\right) \right] \right\}  \nonumber \\
&+& \alpha _2\{Tr\left( \widetilde{\Phi }^{\dagger }\Phi \right) Tr\left(
\Delta _R\Delta _R^{\dagger }\right) +Tr\left( \widetilde{\Phi }\Phi
^{\dagger }\right) Tr\left( \Delta _L\Delta _L^{\dagger }\right)  \nonumber
\\
&+& Tr\left( \widetilde{\Phi }\Phi ^{\dagger }\right) Tr\left( \Delta
_R\Delta _R^{\dagger }\right) +Tr\left( \widetilde{\Phi }^{\dagger }\Phi
\right) Tr\left( \Delta _L\Delta _L^{\dagger }\right) \}  \nonumber \\
&+& \alpha _3\left[ Tr\left( \Phi \Phi ^{\dagger }\Delta _L\Delta _L^{\dagger
}\right) +Tr\left( \Phi ^{\dagger }\Phi \Delta _R\Delta _R^{\dagger }\right)
\right]  \nonumber \\
&+& \beta _1\left[ Tr\left( \Phi \Delta _R\Phi ^{\dagger }\Delta _L^{\dagger
}\right) +Tr\left( \Phi ^{\dagger }\Delta _L\Phi \Delta _R^{\dagger }\right)
\right]  \nonumber \\
&+& \beta _2\left[ Tr\left( \widetilde{\Phi }\Delta _R\Phi ^{\dagger }\Delta
_L^{\dagger }\right) +Tr\left( \widetilde{\Phi }^{\dagger }\Delta _L\Phi
\Delta _R^{\dagger }\right) \right]  \nonumber \\
&+& \beta _3\left[ Tr\left( \Phi \Delta _R\widetilde{\Phi }^{\dagger }\Delta
_L^{\dagger }\right) +Tr\left( \Phi ^{\dagger }\Delta _L\widetilde{\Phi }
\Delta _R^{\dagger }\right) \right] \nonumber\\
&+& \beta _4\left[ Tr\left( \widetilde{\Phi} \Delta _R\widetilde{\Phi }^
{\dagger }\Delta_L^{\dagger }\right) +Tr\left( \widetilde{\Phi} ^{\dagger }
\Delta _L\widetilde{\Phi }\Delta _R^{\dagger }\right) \right]\,,
\end{eqnarray*}
where $\tilde{\Phi}=\tau^2\Phi^*\tau^2$, $\tau^2$ being the Pauli spin 
matrix and $\mu_a^2>0$, with $a=\sigma,\Delta,\Phi_1,\Phi_2$.

\section{Parities in left-right symmetric models and 
consequences}\label{parity}
Now we briefly discuss the parities, $P$ and $D$, in left-right
symmetric models. The main difference between a $D$-parity and
$P$-parity is that the $D$-parity acts on the groups 
$SU(2)_L \otimes SU(2)_R$, while the $P$-parity acts on the
Lorentz group. In the left-right symmetric models we identify
both the parities with each other, so that when we break the
$SU(2)_R$ group or the $D$-parity, the Lorentz $P$-parity is
also broken.

Under the operation of parity the fermions, scalars and the vector 
bosons transform as:
\bea
\psi_{L,R} &\longrightarrow& \psi_{R,L}\nonumber\\
\Phi &\longrightarrow& \Phi^\dagger \nonumber\\
\Delta_{L,R} &\longrightarrow& \Delta_{R,L}\nonumber\\
\sigma &\longrightarrow& -\sigma \nonumber\\
W_{L,R} &\longrightarrow& W_{R,L} .
\label{p-parity}
\eea
This implies that the combinations $W_L + W_R$ and $\Delta_L
+ \Delta_R$ are even under parity, while $W_L - W_R$ and
$\Delta_L - \Delta_R$ are odd under parity. So, $W_L - W_R$ 
is axial vector and $\sigma$ and $\Delta_L - \Delta_R$ are pseudo scalars.
Thus the $vev$ of the fields $\sigma$ or $\Delta_R$ 
can break parity spontaneously. 

It is possible to break the $D$-parity spontaneously by breaking
the group $SU(2)_R$ spontaneously by the $vev$ of the field  $\Delta_R$,
or by breaking it by the $vev$ of $\sigma$. In general, $\sigma$ could
be a scalar or pseudo scalar. If we start with $\sigma$ to be a scalar,
then it can break the $D$-parity keeping the $P$-parity invariant.
However, if we consider $\sigma$ to be a pseudo scalar, it can break
both $D$ and $P$ parities spontaneously. Since it is conventional to
identify $P$ parity with the $D$ parity, we consider $\sigma$ to be
a pseudo scalar. Then the $vev$ of the field $\sigma$ will break parity
and the group $SU(2)_R$ at different scales. This will have some
interesting phenomenology. This was proposed in ref.~\cite{paridaetal.84}. 
Recently its phenomenological consequences using doublet and triplet Higgses 
are studied in ref.~\cite{utpal_plb}.  

We assume that $\mu_\sigma^2>0$ in equation (\ref{potential}). As 
a result below the critical temperature $T_c\sim \langle \sigma \rangle$, 
the parity breaking scale, the singlet Higgs field acquires a $vev$ 
\be
\eta_P \equiv \langle \sigma \rangle= \frac{\mu_\sigma}{\sqrt{
2\lambda_\sigma}}.
\label{sigma-vev}
\ee
Since $\sigma$ doesn't possess any quantum number under $SU(2)_{L,R}$ 
and $U(1)_{B-L}$, these groups remain intact while $P$ breaks. However 
it creates a mass splitting between the triplet fields $\Delta_L$ and 
$\Delta_R$ since it couples differently with them as given in equation 
(\ref{potential}). This leads to different effective masses for $\Delta_L$ 
and $\Delta_R$ 
\bea
M_{\Delta_L}^2 &=& \mu_\Delta^2- (M \eta_P+\gamma \eta_P^2)\,,
\label{deltaL-mass}\\
M_{\Delta_R}^2 &=& \mu_\Delta^2+ (M \eta_P-\gamma \eta_P^2).
\label{deltaR-mass}
\eea
We now apply a fine tuning to set $M_{\Delta_R}^2>0$ so that $\Delta_R$ 
can acquire a $vev$ 
\be
\langle \Delta_{R}\rangle =\begin{pmatrix}
0 & 0\\
v_R & 0 \end{pmatrix}. 
\label{right_vev}
\ee
In order to restore the $SM$ prediction, i.e., to restore the observed 
phenomenology at a low scale, $\Phi$ and $\tilde{\Phi}$ acquire $vev$s 
\be
\langle \Phi \rangle =\begin{pmatrix}
k_1 & 0\\
0 & k_2 \end{pmatrix}~~{\rm and}~~\langle \tilde{\Phi} \rangle =
\begin{pmatrix}k_2 & 0\\
0& k_1 \end{pmatrix}.
\label{dirac_vev}
\ee
This breaks the gauge group $SU(2)_L\otimes SU(2)_R\otimes U(1)_{B-L}$ 
down to $U(1)_{em}$. However, this induces a non-trivial $vev$ for the 
triplet $\Delta_L$ as 
\be
\langle \Delta_{L}\rangle =\begin{pmatrix}
0 & 0\\
v_L & 0 \end{pmatrix}.
\label{left_vev}
\ee
In the above $v_L$, $v_R$, $k_1$ and $k_2$ are real parameters. Further 
the observed phenomenology requires that $v_L\ll k_1,k_2 \ll v_R$. 

Using equations (\ref{right_vev}), (\ref{dirac_vev}) and 
(\ref{left_vev}) in equation (\ref{potential}) we get the effective 
potential 
\begin{eqnarray}
\mathbf{V}_{eff} &=&-\mu_\sigma^2 \eta_P^2\nonumber\\
&&-\left[ \mu_\Delta^2-M\eta_P-\gamma\eta_P^2 -\alpha_{1}(k_1^2+k_2^2)
-\alpha_{2}(4k_1k_2)-\alpha_{3}k_2^2\right]v_L^2\nonumber\\ 
&&-\left[ \mu_\Delta^2 + M\eta_P-\gamma\eta_P^2-\alpha_{1}(k_1^2+k_2^2)
-\alpha_{2}(4k_1k_2)-\alpha_{3}k_2^2\right]v_R^2\nonumber\\
&&-\mu_{\Phi 1}^2 (k_1^2+k_2^2)-\mu_{\Phi 2}^2(4k_1k_2)\nonumber\\
&&+\lambda_\sigma\eta_P^4+\rho_1(v_L^4+v_R^4)+\rho_3v_L^2v_R^2\nonumber\\
&&+\lambda_1(k_1^2+k_2^2)+(2\lambda_2+\lambda_3)(4k_1^2k_2^2)
+\lambda_4(k_1^2+k_2^2)(4k_1k_2)\nonumber\\
&&+\delta_1\eta_P^2(k_1^2+k_2^2)+\delta_2\eta_P^2(4k_1k_2)\nonumber\\
&&+2(\beta_1k_1k_2+\beta_2k_1^2+\beta_3k_2^2+\beta_4k_1k_2)v_Lv_R.
\label{eff-pot}
\end{eqnarray}
The electroweak phase transition occurs at a low energy scale and 
hence it is reasonable to assume that the parameters $k_2^2, 
k_1k_2, k_1^2\ll \eta_P$. Using this approximation in equation 
(\ref{eff-pot}) one can see that the effective masses of 
$\Delta_L$ and $\Delta_R$ coincides with equations (\ref{deltaL-mass}) 
and (\ref{deltaR-mass}). Further assuming $M=\gamma \eta_P$ we get
\be
M_{\Delta_R}^2=\mu_\Delta^2~~{\rm and}~~M_{\Delta_L}^2=
M_{\Delta_R}^2-2 \gamma\eta_P^2.
\label{cancellation}
\ee
Thus a large cancellation between $M_{\Delta_R}$ and $\gamma\eta_P$ 
allows an effective mass scale of the triplet $\Delta_L$ to be 
very low and vice-versa. 

We now check the order of magnitude of the induced $vev$ of the 
triplet $\Delta_L$. This should be small (less than a GeV) in 
order the theory to be consistent with $Z$-decay width. Further 
the sub-eV masses of the light neutrinos require $vev$ of $\Delta_L$ 
to be of the order of eV, because it gives masses through the 
type-II seesaw mechanism. From equation (\ref{eff-pot}) we get 
\begin{eqnarray}
v_R\frac{\partial \mathbf{V}_{eff}}{\partial v_L}- v_L 
\frac{\partial \mathbf{V}_{eff}}{\partial v_R} &=& 0\nonumber\\
&=& v_Lv_R[4M\eta_P-4\rho_1(v_R^2-v_L^2)+2\rho_3(v_R^2-v_L^2)]\nonumber\\
&+& 2(\beta_1k_1k_2+\beta_2k_1^2+\beta_3k_2^2+\beta_4k_1k_2)(v_R^2-v_L^2).
\label{minimisation}
\end{eqnarray}
In the quark sector the $vev$s $k_1$ and $k_2$ give masses to the up 
and down type quarks respectively. Therefore, it is reasonable to assume 
\be
\frac{k_1}{k_2}=\frac{m_t}{m_b}\,.
\label{top-bottom-ratio}
\ee
With the approximation $v_L\ll k_1,k_2\ll v_R\ll \eta_P$ and using the 
above assumption (\ref{top-bottom-ratio}) in equation (\ref{minimisation}) 
we get 
\be
v_L\simeq \frac{-\beta_2 v^2v_R}{2M\eta_P}\,,
\label{vev_value}
\ee
where we have used $v=\sqrt{k_1^2+k_2^2}\simeq k_1=174$ GeV.
Notice that in the above equation the smallness of the $vev$ of 
$\Delta_L$ is decided by the parity breaking scale, not the 
$SU(2)_R$ breaking scale. So there are no constraints on 
$v_R$ from the seesaw point of view. After $SU(2)_R$ symmetry 
breaking the right handed neutrinos acquire masses through the Majorana 
Yukawa coupling with the $\Delta_R$. Depending on the strength of 
Majorana Yukawa coupling a possibility of TeV scale right handed 
neutrino is unavoidable. We will discuss the consequences in context 
of $L$-asymmetry in section IV. 

Finally before going to discuss the $L$-asymmetry in this model 
we give a most economic breaking scheme of $SO(10)$ GUT through the 
left right symmetric path. Keeping in mind that the $P$ and 
$SU(2)_R$ breaking scales are different, the breaking of $SO(10)$ 
down to $U(1)_{em}$ can be accomplished by using a set of Higgses: 
$\{210\}$, $\{126\}$, $\{10\}$ of $SO(10)$. At the first stage $SO(10)$ 
breaks to $G_{224}\equiv SU(2)_L\otimes SU(2)_R\otimes SU(4)_C (
\supset SU(2)_L\otimes SU(2)_R\otimes U(1)_{B-L}\otimes SU(3)_C)$ through 
the $vev$ of $\{210\}$. Under $G_{224}$ its decomposition can be 
written as 
\be
\{210\}=(1,1,1)+(2,2,20)+(3,1,15)+(1,3,15)+(2,2,6)+(1,1,15)\,,
\label{210-higgs}
\ee
where $(1,1,1)$ is a singlet and it is odd under the $D$ parity,
which is a generator of the group $SO(10)$. Hence it 
can play the same role as $\sigma$ 
discussed above. At a later epoch \{126\} of $SO(10)$ can get a $vev$ and 
breaks $SU(2)_L\otimes SU(2)_R\otimes SU(4)_C$ to $G_{213}\equiv 
SU(2)_L\otimes U(1)_Y\otimes SU(3)_C$. Under $G_{224}$ the decomposition 
of $\{126\}$ is given as 
\be
\{126\}=(3,1,10)+(1,3,10)+(2,2,15)+(1,1,6)\,,
\label{126-higgs}
\ee
where $(3,1,10)$ and $(1,3,10)$ contain the fields $\Delta_L$ 
and $\Delta_R$ respectively as in the above discussion. Finally the 
$vev$ of $\{10\}$ breaks the gauge group $SU(2)_L\otimes U(1)_Y
\times SU(3)_C $ down to $U(1)_{em}\otimes SU(3)_C$ which contains 
a $(2,2,1)$ playing the role of $\Phi$ in our discussion.

\section{Neutrino masses and leptogenesis in left-right symmetric models}
The relevant Yukawa couplings giving masses to the three generations 
of leptons are given by
\bea
\mathcal{L}_{yuk} &=& h_{ij}\overline{\psi_{Li}}\psi_{Rj}\Phi+
\tilde{h}_{ij}\overline{\psi_{Li}}\psi_{Rj}\tilde{\Phi}+ H.C.\nonumber\\
&+& f_{ij}\left[\overline{(\psi_{Li})^c}\psi_{Lj}\Delta_L+
\overline{(\psi_{Ri})^c}\psi_{Rj}\Delta_R\right]+ H.C.\,,
\label{yukaw}
\eea
where $\psi_{L,R}^T=(\nu_{L,R}, e_{L,R})$. The discrete left-right 
symmetry ensures the Majorana Yukawa coupling $f$ to be same for 
both left and right handed neutrinos. The breaking of left-right 
symmetry down to $U(1)_{em}$ results in the effective mass matrix 
of the light neutrinos to be 
\bea
m_\nu &=& fv_L - m_D\frac{f^{-1}}{v_R}m_D^T\nonumber\\
&=& m_\nu^{II}+m_\nu^I\,,
\label{neutrino-mass}
\eea
where $m_D=h k_1+\tilde{h} k_2\simeq h k_1$ and $v_L$ is given by equation 
(\ref{vev_value}). In theories where both type-I and type-II mass 
terms originate at the same scale it is difficult to choose 
which of them contribute dominantly to the neutrino mass matrix. 
In contrast to it in the present case since the parity and the 
$SU(2)_R$ breaking scales are different and, in fact, $\eta_P\gg v_R$ 
it is reasonable to assume that the type-I neutrino mass dominantly 
contributes to the effective neutrino mass matrix. In what follows 
we assume 
\be
m_\nu=m_\nu^I=- m_D\frac{f^{-1}}{v_R}m_D^T.
\label{typeI-mass}
\ee

In the previous section we showed that the $SU(2)_R$ breaking 
scale $v_R$ can be much lower than the parity breaking scale 
$\eta_P$ since the smallness of $v_L$ doesn't depend on $v_R$. 
Conventionally this leads to the right handed neutrino masses 
to be smaller than that of the triplet $\Delta_L$~\cite{paridaetal.84}. 
However, in the present case a large cancellation between 
$M_{\Delta_R}^2$ and $\gamma \eta_P^2$ allows an 
effective mass of the triplet $\Delta_L$ to be in low scale while 
leaving the mass of $\Delta_R$ at the $D$-parity breaking 
scale. Note that the source of smallness of the right handed neutrinos 
and the triplet $\Delta_L$ are absolutely different. Unless the low 
energy observables constrain their masses one can't predict which one 
is lighter. In the following we take leptogenesis as a tool 
to distinguish their mass scales.   
    
\subsection{Leptogenesis via heavy neutrino decay}
\begin{figure}[htbp]
\begin{center}
\epsfig{file=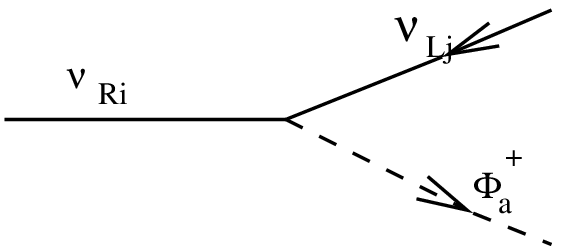, width=0.3\textwidth}
\epsfig{file=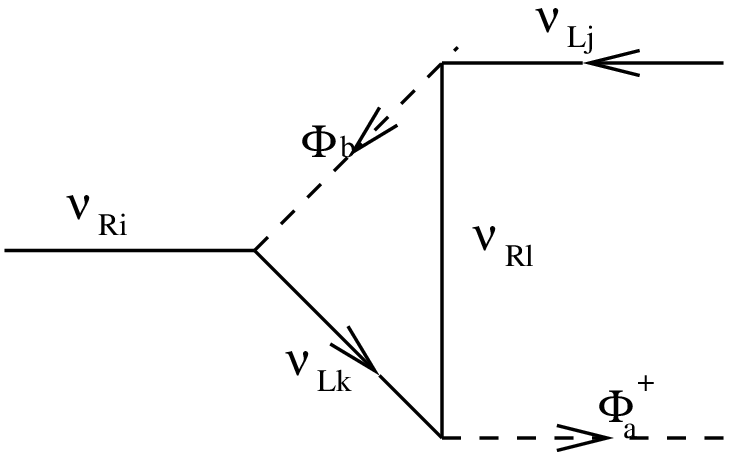, width=0.3\textwidth}
\epsfig{file=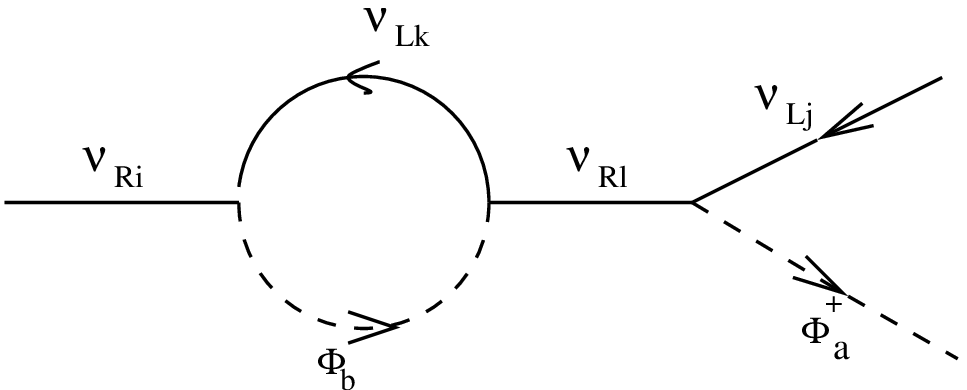, width=0.35\textwidth}
\caption{The tree level, one loop radiative correction and the 
self energy correction diagrams contributing to the $CP$-asymmetry 
in the decay of heavy Majorana neutrinos.}
\label{maj-asy}
\end{center}
\end{figure}

Without loss of generality we work in a basis in which the mass 
matrix of the right handed neutrinos is real and diagonal. In 
this basis the heavy Majorana neutrinos are defined as 
$N_i=(1/\sqrt{2})(\nu_{Ri}\pm \nu_{Ri}^c)$, where i=1,2,3 representing 
the flavor indices. The corresponding masses of the heavy Majorana 
neutrinos are given by $M_i$. In this basis a net $CP$-asymmetry 
results from the decay of $N_i$ to the $SM$ fermions and the 
bidoublet Higgses and is given by the interference of tree level, 
one loop radiative correction and the self-energy correction diagrams 
as shown in figs.(\ref{maj-asy}).
The resulting $CP$-asymmetry in this case is given by 
\be
\epsilon_i^I=\frac{1}{8\pi}\frac{\sum_{l} Im\left[ ({h^a}^\dagger h^b)_
{il} ({h^b}^\dagger h^a)_{il} \right]} {({h^a}^\dagger h^a)_{ii}}\sqrt{x_l}
\left[ 1-(1+x_l)\log (1+1/x_l)+1/(1-x_l) \right]\,,
\label{majorana-cp}
\ee
where $x_l=M_l^2/M_i^2$ and $h^a$, with $a=1,2$ stands for the Dirac 
Yukawa couplings of fermions with $\Phi$ and $\tilde{\Phi}$ respectively. 
That is $h^1=h$ and $h^2=\tilde{h}$ as given in equation (\ref{yukaw}). 
Now we assume a normal mass hierarchy, $M_1\ll M_2<M_3$, in the heavy 
Majorana neutrino sector. In this case while the heavier right handed 
neutrinos $N_2$ and $N_3$ are decaying yet the lightest one, $N_1$, is 
in thermal equilibrium. Any $L$-asymmetry thus produced by the decay 
$N_2$ and $N_3$ is erased by the $L$-number violating scatterings 
mediated by $N_1$. Therefore, it is reasonable to assume that the 
final $L$-asymmetry is given by the decay of $N_1$. Simplifying 
equation (\ref{majorana-cp}) we get a net $CP$-asymmetry coming from 
the decay of $N_1$ to be  
\be
\epsilon_1^{I}=-\frac{3M_1}{16\pi }\frac{\sum_{i,j}Im
\left[ ({h^a}^\dagger)_{1i} (h^b (M_{dia})^{-1}(h^a)^T)_{ij}
({h^b}^*)_{j1}\right]}{({h^a}^\dagger h^a)_{11}}\,.
\label{epsilonI}
\ee
Expanding the above equation (\ref{epsilonI}) and using the fact 
that $m_\nu \simeq -k_1^2(hM_{dia}^{-1}h^T)$ we get 
\be
\epsilon_1^{I}=\frac{3M_1}{16\pi v^2}\left\{ \frac{\sum_{i,j}Im
\left[ ({h}^\dagger)_{1i} (m_{\nu}^I)_{ij}({h}^*)_{j1}\right]}
{({h^a}^\dagger h^a)_{11}} + (h, \tilde{h}) {\rm terms}\right\}.
\label{typeI-epsilon}
\ee
Unlike the type-I models~\cite{type-I-bound} here we have additional terms 
contributing the $CP$-asymmetry in the decay of $N_1$. Note that if the 
strength of $\tilde{h}$ is comparable with $h$ then the resulting 
$CP$-asymmetry enhances by a factor of 2 in comparison with the 
$CP$-asymmetry in the exclusive type-I models~\cite{type-I-bound}. 

\begin{figure}[htbp]
\begin{center}
\epsfig{file= 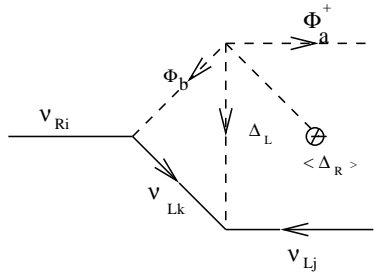, width=0.3\textwidth}
\caption{The one loop radiative correction through the virtual 
triplet $\Delta_L$ in the decay of right handed heavy Majorana 
neutrino contributes to the $CP$-asymmetry.}
\label{triplet-correction}
\end{center}
\end{figure}

An additional contribution to $CP$-asymmetry also comes from the interference 
of tree level diagram in fig. (\ref{maj-asy}) and the one loop radiative 
correction diagram involving the virtual triplet $\Delta_L$ as shown in 
fig. (\ref{triplet-correction}).
The resulting $CP$-asymmetry in this case is given by~\cite{hamby&sen_plb,
triplet-cp-asy}
\be
\epsilon_i^{II}=\frac{3}{8\pi}\frac {\sum_{k,j}Im \left[ (h^a)_{ji}^* 
(h^b)_{ki}^* f_{jk}(v_R\beta)_{ab}\right]}{(h^a{h^a}^\dagger)_{ii}M_i} 
\left(1-\frac{M_{\Delta_L}^2}{M_i^2}\log (1+M_i^2/M_{\Delta_L}^2) \right)\,,
\label{maj-tri-asy}
\ee
where 
\be
\beta=\begin{pmatrix}\beta_1&\beta_3\\
\beta_2 & \beta_4\end{pmatrix}\,.
\label{beta_values}
\ee
If we further assume that $M_1\ll M_{\Delta_L}$ in addition to the 
normal mass hierarchy in the heavy Majorana neutrino sector, then the 
final $L$-asymmetry must be given by the $CP$-violating decay of 
$N_1$ to the $SM$ lepton and the bidoublet Higgs. Now using 
(\ref{vev_value}) in equation (\ref{maj-tri-asy}) we get the 
$CP$-asymmetry parameter
\be
\epsilon_1^{II}= \frac{3M_1}{16\pi v^2} \left( \frac{2M\eta_P}{-\beta_2 
M_{\Delta_L}^2} \right) \frac{ \sum_{jk}Im \left[ ({h^a}^{\dagger})_{1j} 
(m_\nu^{II})_{jk}(h^b)^*_{k1} \beta_{ab}\right]}{({h^a}^\dagger 
h^a)_{11}}.
\label{epsilonII}
\ee
Note that this result differs from the usual type-II seesaw models~\cite{
hamby&sen_plb,antusch&king_plb} where only one triplet $\Delta_L$ is 
usually introduced into the $SM$ in addition to the singlet heavy Majorana 
neutrinos. 

The total $CP$-asymmetry coming from the decay of $N_1$ thus reads
\be
\epsilon_1=\epsilon_1^I+\epsilon_1^{II}\,,
\label{epsilon}
\ee
where $\epsilon_1^I$ and $\epsilon_1^{II}$ are given by equations 
(\ref{typeI-epsilon}) and (\ref{epsilonII}) respectively. Unlike the 
existing literature~\cite{antusch&king_plb,type-II-bound} in the 
present case it is impossible to compare the magnitude of $\epsilon_1^I$ 
and $\epsilon_1^{II}$ through the type-I and type-II neutrino mass terms 
unless one takes the limiting cases. 

\subsubsection{Dominating type-I contribution}\label{dom-type-I}
Let us first assume that $\epsilon_1^I$ dominates in equation 
(\ref{epsilon}) and the neutrino Dirac Yukawa coupling 
$h\simeq \tilde{h}$. The resulting $CP$-asymmetry is then given by 
\be
\epsilon_1=\epsilon_1^{I}=2\left\{ \frac{3M_1}{16\pi v^2}\frac{\sum_{i,j}Im
\left[ ({h}^\dagger)_{1i} (m_{\nu}^I)_{ij}({h}^*)_{j1}\right]}
{({h}^\dagger h)_{11}} \right\}\,.
\label{resulting-epsilon}
\ee
The maximum value of $\epsilon_1$ then reads $\epsilon_1^{max}=
2\epsilon_1^{0}$~\cite{type-I-bound}, where 
\be
|\epsilon_1^{0}|=\frac{3M_1}{16\pi v^2}\sqrt{\Delta m_{atm}^2}\,.
\label{cp-max-I}
\ee
As a result we gain a factor of 2 in the lower bound on $M_1$ 
which is given as 
\begin{equation}
M_1\geq 4.2\times 10^8 GeV \left(\frac{n_B/n_\gamma}{6.4\times 10^{-10}}
\right) \left(\frac{10^{-3}}{\frac{n_{\nu_R}}{s}
\delta}\right)\left(\frac{v}{174GeV}\right)^2\left(\frac{0.05eV}
{\sqrt{\Delta m_{atm}^2}}\right)\,,
\label{lower-bound-M1}
\end{equation}
where we have made use of the equation (\ref{asymmetry}). 
 
\subsubsection{Dominating type-II contribution}
Suppose $\epsilon_1^{II}$ dominates in equation (\ref{epsilon}).
In that case, assuming $\tilde{h}\simeq h$ and $\beta_i$'s of order 
unity we get the maximum value of the $CP$-asymmetry parameter~\cite{
type-II-bound} 
\be
|\epsilon_1^{max}|=\left( \frac{4M\eta_P}{M_{\Delta_L}^2} \right) 
\frac{3M_1}{16\pi v^2}m_3\,,
\label{cp-max-II}
\ee
where $m_3=\sqrt{\Delta m_{atm}^2}\simeq 0.05$ eV. Following the 
same procedure in section (\ref{dom-type-I}) we gain a factor 
of ($M_{\Delta_L}^2/4M\eta_P$) in the lower bound on $M_1$.

\subsection{Leptogenesis through triplet decay}
In the left-right symmetric models the decay of the scalar 
triplets $\Delta_L$ and $\Delta_R$ violates $L$-number by 
two units and hence potentially able to produce a net $L$-asymmetry. 
The efficient decay modes which violate $L$-number are  
\bea
\Delta_{L,R} &\longrightarrow & \nu_{L,R}+\nu_{L,R}\,,\nonumber\\
\Delta_{L,R} &\longrightarrow & {\Phi^a}^\dagger +\Phi^b .
\label{decay-modes}
\eea
However, the decay rate in the process $\Delta_R\longrightarrow 
{\Phi^a}^\dagger+ \Phi^b $ is highly suppressed in comparison to 
$\Delta_L \longrightarrow {\Phi^a}^\dagger+\Phi^b$ because of the 
proportionality constant is $v_L$ in the former case while it is 
of $v_R$ in the latter case. Moreover, in the present case the 
effective mass scale of the triplet $\Delta_R$ is larger than the 
mass of $\Delta_L$ due to the large cancellation between 
$M_{\Delta_R}^2$ and $2\gamma\eta_P^2$. Therefore, in what follows 
we take only the decay modes of the triplet $\Delta_L$. The decay 
rates are given as: 
\bea
\Gamma_\nu(\Delta_L\rightarrow\nu_{Li}\nu_{Lj}) &=& 
\frac{|f_{ij}|^2}{8\pi}M_{\Delta_L}\label{tri-lep-decay}\,,\\
\Gamma_\Phi(\Delta_L\rightarrow {\Phi^a}^\dagger\Phi^b) &=& 
\frac{|\beta_{ab}|^2}{8\pi}r^2 M_{\Delta_L}\,,
\label{tri-phi-decay}
\eea
where $\beta_{ab}$ are given in equation (\ref{beta_values}) and 
$r^2=(v_R^2/M_{\Delta_L}^2)$. A net asymmetry is produced when 
the decay rate of the triplet $\Delta_L$ fails to compete with the 
Hubble expansion rate of the Universe. This is given by the conditions:  
\bea
\Gamma_\nu &\lsim & H(T=M_{\Delta_L})\label{tri-lep-rate}\,,\\
\Gamma_\Phi &\lsim & H(T=M_{\Delta_L})\label{tri-phi-rate}.
\eea

As shown in equation (\ref{cancellation}) a large cancellation 
can lead to a TeV scale of the triplet $\Delta_L$. However, the 
SM gauge interaction $W_L^\dagger +W_L\longrightarrow\Delta_L^\dagger +
\Delta_L$ keeps it in thermal equilibrium. The out of equilibrium of 
this process requires $\Gamma_W\leq H(T=M_{\Delta_L})$. Consequently 
we will get a lower bound on the mass of the triplet $\Delta_L$ to 
be $M_{\Delta_L}\geq 4.8\times 10^{10}$ GeV. 

\begin{figure}[htbp]
\begin{center}
\epsfig{file=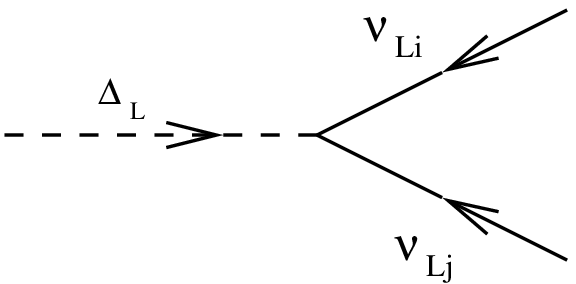, width=0.3\textwidth}
\epsfig{file=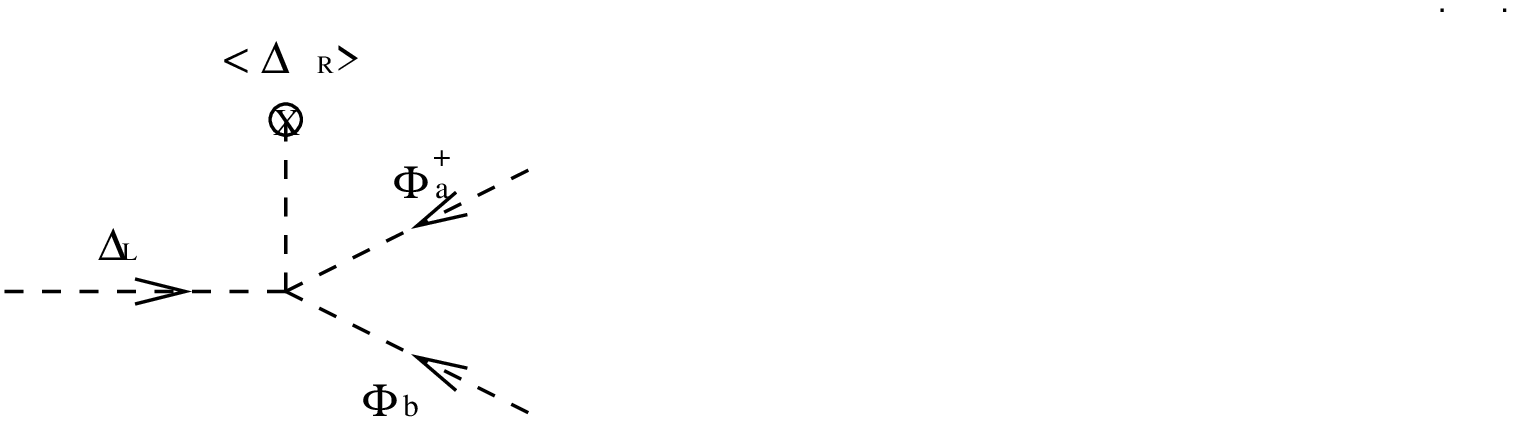, width=0.3\textwidth}
\caption{The tree level diagrams of the decay of the triplet 
$\Delta_L$ contributing to the $CP$-asymmetry.}
\label{tripletasymmetry}
\end{center}
\end{figure}

\begin{figure}
\begin{center}
\epsfig{file=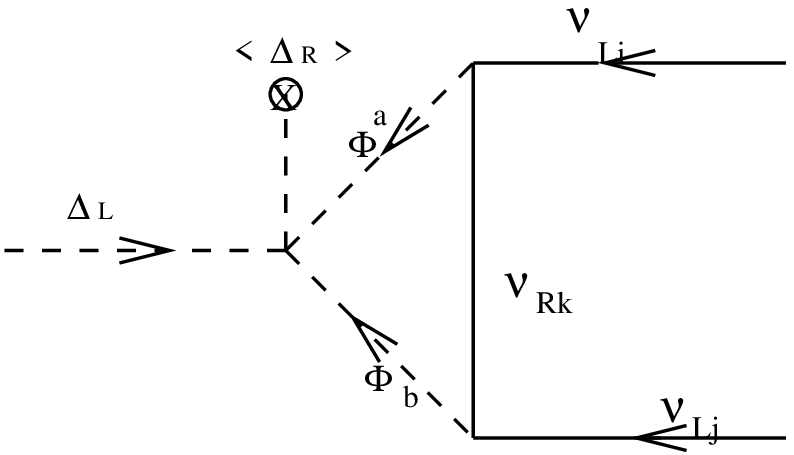, width=0.3\textwidth}
\epsfig{file=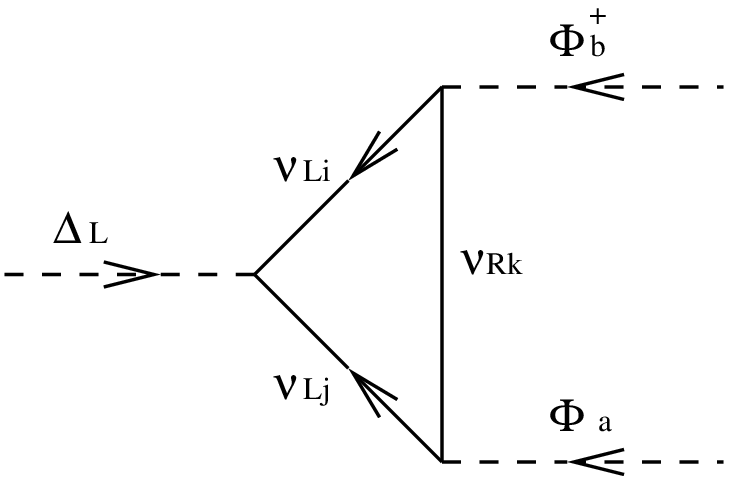, width=0.3\textwidth}
\caption{The one loop radiative correction of the decay of the 
$\Delta_L$ through the exchange of virtual right handed neutrinos 
contributing to the $CP$-asymmetry.}
\label{triplet-loop-asymmetry}
\end{center}
\end{figure}

The $CP$-asymmetry in this case arises from the interference of tree level 
diagrams in figs. (\ref{tripletasymmetry}) with the one loop radiative 
correction diagrams involving the virtual right handed neutrinos as shown 
in the figs: (\ref{triplet-loop-asymmetry}).
The resulting $CP$-asymmetry in this case is given by~\cite{hamby&sen_plb,
triplet-cp-asy} 
\be
\epsilon_{\Delta}=\frac{1}{8\pi}\sum_k M_k \frac{\sum_{ij}Im 
\left[ (h^a)^*_{ik}(h^b)^*_{jk}(\beta v_R)_{ab}^* f_{ij}\right]}
{\sum_{ij}|f_{ij}|^2 M_{\Delta_L}^2+\sum_{cd}|\beta_{cd}|^2v_R^2}
\log (1+\frac{M_{\Delta_L}^2}{M_k^2}).
\label{triplet-asymmetry}
\ee
Assuming that $M_{\Delta_L}<M_1$ and $h=\tilde{h}$ the above 
equation can be simplified to  
\be
\epsilon_{\Delta}=\frac{1}{8\pi v^2}\frac{\sum_{ij}Im 
\left[ (m_\nu^I)^*_{ij}(M_R)_{ij\sum \beta^*}\right]}{\sum_{ij}|f_{ij}|^2
+ \sum_{cd}|\beta_{cd}|^2r^2}\,,
\label{triplet-asy}
\ee 
where $m_\nu^I$ is given by equation (\ref{typeI-mass}) which can be 
calculated from the low energy neutrino oscillation data. 

\section{charge-neutral symmetry and neutrino mass matrices}
The present neutrino oscillation data show that the neutrino mixing
matrix up to a leading order in $\sin\theta_{13}$ is~\cite{pmns}
\be
U_{PMNS}=\begin{pmatrix}\frac{\sqrt{2}}{\sqrt{3}}& \frac{1}{\sqrt{3}}&
\epsilon e^{-i\delta}\\
\frac{-1}{\sqrt{6}}-\frac{1}{\sqrt{3}}\epsilon e^{i\delta} & 
\frac{1}{\sqrt{3}}-\frac{-1}{\sqrt{6}}\epsilon e^{i\delta} & 
\frac{1}{\sqrt{2}}\\
\frac{1}{\sqrt{6}}-\frac{1}{\sqrt{3}}\epsilon e^{i\delta} & 
\frac{-1}{\sqrt{3}}-\frac{-1}{\sqrt{6}}\epsilon e^{i\delta} & 
\frac{1}{\sqrt{2}}\end{pmatrix}{\mathrm dia}\left(1, e^{i\alpha},
e^{(\beta+\delta)}\right)\,
\label{pmns-matrix}
\ee
where we have used the best fit parameters~\cite{neu-parameter}; the
atmospheric mixing angle $\theta_{23}=45^\circ$, the solar mixing
angle $\theta_{12}\simeq 34^\circ$ and the reactor angle $\sin \theta_{13}
\equiv \epsilon$. Using (\ref{pmns-matrix}) the neutrino mass matrix
can be written as
\be
m_\nu=U_{PMNS}^* m_\nu^{dia} U_{PMNS}^\dagger\,,
\label{neu-mass}
\ee
where $m_\nu^{dia}=dia(m_1,m_2,m_3)$, with $m_1, m_2, m_3$ are the
light neutrino masses. Using equations (\ref{pmns-matrix}) and 
(\ref{neu-mass}) we get, up to an order of $\epsilon$, the elements 
of the neutrino mass matrix:
\bea
(m_\nu)_{11} &=& \frac{m_2}{3}+\frac{2}{3}m_1\nonumber\\
(m_\nu)_{12} &=& \epsilon e^{i\delta}\frac{m_3}{\sqrt{2}} + \frac{m_2}{
\sqrt{3}} \left( \frac{1}{\sqrt{3}}-\frac{1}{\sqrt{6}}\epsilon 
e^{-i\delta}\right) - \sqrt{{2\over 3}}m_1 \left( \frac{1}{\sqrt{6}}+
\frac{\epsilon e^{-i\delta}}{\sqrt{3}} \right) \nonumber\\
(m_\nu)_{13} &=& \epsilon e^{i\delta} \frac{m_3}{\sqrt{2}}-\frac{m_2}
{\sqrt{3}}\left( \frac{1}{\sqrt{3}}+\frac{1}{\sqrt{6}}\epsilon
e^{-i\delta} \right) +\sqrt{{2\over 3}}m_1 \left( \frac{1}{\sqrt{6}}-
\frac{\epsilon e^{-i\delta}}{\sqrt{3}} \right) \nonumber\\
(m_\nu)_{23} &=& \frac{m_3}{2}-\frac{m_2}{3}-\frac{m_1}{6}\nonumber\\
(m_\nu)_{22} &=& \frac{m_3}{2}+\frac{m_1}{\sqrt{6}} \left(\frac{1}{\sqrt{6}}
+\frac{2\epsilon e^{-i\delta}}{\sqrt{3}} \right)+\frac{m_2}{\sqrt{3}}
\left( \frac{1}{\sqrt{3}}-\sqrt{\frac{2}{3}}\epsilon e^{-i\delta} 
\right)\nonumber\\
(m_\nu)_{33} &=& \frac{m_3}{2}+ \frac{m_1}{\sqrt{6}} \left(\frac{1}{\sqrt{6}}
-\frac{2\epsilon e^{-i\delta}}{\sqrt{3}} \right)+\frac{m_2}{\sqrt{3}}
\left( \frac{1}{\sqrt{3}}+ \frac{2\epsilon e^{-i\delta}}{\sqrt{6}} \right)
\label{neu-masses}
\eea

Inverting the seesaw relation (\ref{typeI-mass}) we get the right 
handed neutrino mass matrix~\cite{inverse-seesaw}
\be
M_R=-m_D^T m_\nu^{-1}m_D\,,
\label{Maj-mass-matrix}
\ee
where $M_R=fv_R$. The $m_\nu^{-1}$ in the above equation can be calculated
from equation (\ref{neu-mass}). Unless one assumes a texture of $m_D$ it is
difficult to link $m_\nu$ and $M_R$ through equation (\ref{Maj-mass-matrix}).
In general it is almost impossible to connect the low energy $CP$-phase
and the $CP$-phase appearing in leptogenesis. So, by using some approximations
for the neutrino Dirac mass matrix one can calculate the right handed neutrino
mass matrix $M_R$ and hence the $CP$-asymmetry~\cite{leptogenesis-group}. 
We assume a charge neutral symmetry which is natural in the 
supersymmetric left-right symmetric models~\cite{babu_group}. 
We take the neutrino Dirac mass
\be
m_D=c m_l\,,
\label{neu_dirac_mass}
\ee
where $m_l$ is the charged lepton mass matrix and $c$ is a numerical
factor. Further we assume the texture of the charged leptons mass 
matrix as~\cite{lepton-texture}
\be
m_l=\begin{pmatrix}0&\sqrt{m_em_\mu}&0\\
\sqrt{m_em_\mu}& m_\mu & \sqrt{m_em_\tau}\\
0 & \sqrt{m_em_\tau}&m_\tau \end{pmatrix}.
\label{lepton-mass}
\ee

We shall further assume that at a high energy scale, where the 
leptogenesis occurs, the PMNS matrix is given by~\cite{giunti&tanimoto}
\be
U_{PMNS}=U_l^\dagger U_0\,,
\label{pmns-general}
\ee 
where $U_l$ and $U_0$ are the diagonalizing matrix of $m_l$ and $m_\nu$ 
respectively. At this scale we assume $U_l=I$ and a bimaximal structure 
for $U_0$ which is given by
\be
U_0=\begin{pmatrix}\frac{1}{\sqrt{2}}& \frac{1}{\sqrt{2}}&
\epsilon e^{-i\delta}\\
\frac{-1}{2} & \frac{1}{2} & \frac{1}{\sqrt{2}}\\
\frac{1}{2} & \frac{-1}{2} & \frac{1}{\sqrt{2}}
\end{pmatrix}\,.
\label{bimaximal-form}
\ee
Now using (\ref{neu_dirac_mass}) and (\ref{lepton-mass}) in equation 
(\ref{Maj-mass-matrix}) we get the elements in the right handed neutrino 
mass matrix as:
\bea
(M_R)_{11} &\simeq& -c^2 (m_e m_\mu)\left( \frac{1}{4m_1}(1+2\epsilon 
e^{i\delta}) + \frac{1}{4m_2}(1-2\epsilon e^{i\delta})+
\frac{1}{2m_3} \right)\nonumber\\ 
(M_R)_{12} &\simeq& -c^2 (m_\mu\sqrt{m_em_\mu}) \left(
\frac{1}{4m_1}(1+2\epsilon e^{i\delta}) +\frac{1}{4m_2}
(1-2\epsilon e^{i\delta})+\frac{1}{2m_3} \right)\nonumber\\
(M_R)_{13} & \simeq& -c^2(m_\tau \sqrt{m_e m_\mu}) \left( -\frac{1}{4m_1}
-\frac{1}{4m_2}+\frac{1}{2m_3} \right)\nonumber\\
(M_R)_{22} &\simeq& -c^2 m_\mu^2 \left( \frac{1}{4m_1}(1+2\epsilon 
e^{i\delta}) + \frac{1}{4m_2}(1-2\epsilon e^{i\delta}) + 
\frac{1}{2m_3} \right)\nonumber\\ 
(M_R)_{23} &\simeq& -c^2 (m_\mu m_\tau)\left(-\frac{1}{4m_1}-\frac{1}{4m_2}
+\frac{1}{2m_3} \right)\nonumber\\
(M_R)_{33} &\simeq& -c^2m_\tau^2 \left(\frac{1}{4m_1}(1+2\epsilon
e^{i\delta})+\frac{1}{4m_2}(1-2\epsilon e^{i\delta})+\frac{1}{2m_3}\right).
\label{maj-mass-matrix}
\eea
Below the electroweak phase transition the charged leptons are massive 
and the corresponding mass matrix is given by equation (\ref{lepton-mass}). 
So we can recover the PMNS matrix at low energy as given by equation 
(\ref{pmns-general}) by attributing the small deviation from its bimaximal 
form to the diagonalizing matrix of the charged leptons 
$U_l$~\cite{giunti&tanimoto}.

\section{Lepton asymmetry with charge-neutral symmetry}
In this section we estimate the $L$-asymmetry from the decay 
of right handed neutrino as well as the triplet $\Delta_L$, 
depending on the relative masses they acquire from the symmetry 
breaking pattern.

\subsection{$L$-asymmetry with $M_1<M_{\Delta_L}$ and dominating 
$\epsilon_1^I$}\label{dominate-typeI}
Using (\ref{neu-masses}) and (\ref{neu_dirac_mass}) in equation 
(\ref{resulting-epsilon}) we get the resulting $CP$-asymmetry parameter 
from the decay of right handed neutrino to be 
\be
\epsilon_1^I\simeq -\frac{M_1}{16\pi v^2}\left[ (2m_1+m_2)\epsilon^2 
\sin 2\delta +2\sqrt{2}(m_1-m_2)\epsilon \sin \delta \right]\,.
\label{final-cp-asy}
\ee
The $L$-asymmetry in a comoving volume is then given by 
\be
Y_L=\epsilon_1^I Y_{N_1} d\,,
\label{L-asym}
\ee
where $Y_{N_1}=(n_{N_1}/s)$, $s=(2\pi^2/45)g_*T^3$ is the entropy 
density, $n_{N_1}$ is the number density of lightest right handed 
neutrino in a physical volume and $d$ is the dilution factor which can be 
obtained by solving the required Boltzmann equations. A part of 
the $L$-asymmetry is then transferred to the $B$-asymmetry in a 
calculable way. As a result we get the net $B$-asymmetry 
\be
\frac{n_B}{n_\gamma}=7Y_B=-3.5\epsilon_1^I Y_{N_1} d\,. 
\ee 
\begin{figure}[htbp]
\begin{center}
\epsfig{file=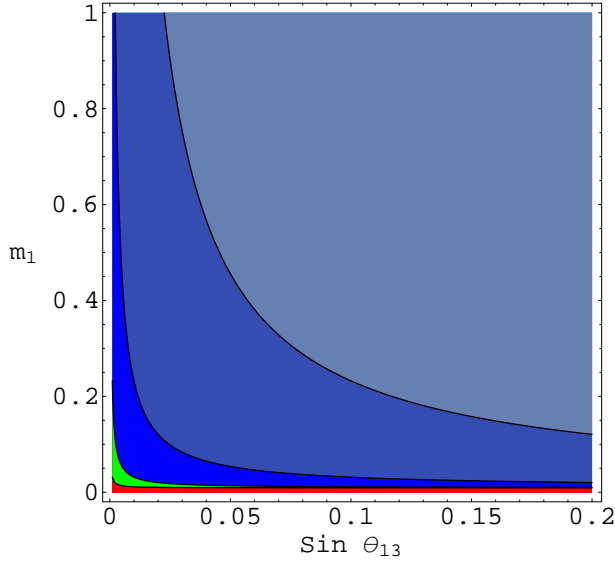}
\caption{Contours satisfying the required B-asymmetry are plotted in 
the $\sin \theta_{13}$ versus $m_1$ plane for $(4.2\times 10^8 
GeV/M_1)=0.1,0.01,0.001,0.0001$} 
\label{typeI-m1_theta13}
\end{center}
\end{figure}

With the maximal CP asymmetry, i.e., $\delta=\pi/2$, and using the 
best fit parameter for $m_2=0.009$ eV we have shown the regions in 
the $\sin \theta_{13}$ versus $m_1$ plane for various values of $M_1$ 
as shown in fig. (\ref{typeI-m1_theta13}). The upper most region 
represents $4.2\times 10^8 GeV$ $< M_1 <$ $4.2\times 10^9$ GeV. As we go 
down the mass of $N_1$ increases by an order of magnitude per region. If we 
assume a normal mass hierarchy for the light physical neutrinos then 
only the bottom most region i.e., $M_1>4.2\times 10^{12}$ GeV, is 
allowed for all $m_1<0.001 eV$ and $\sin \theta_{13}<0.2$, the 
present experimentally allowed values.   

\subsection{$L$-asymmetry with $M_1<M_{\Delta_L}$ and dominating 
$\epsilon_1^{II}$}
Assuming a normal mass hierarchy in the right handed neutrino sector 
and the mass of lightest right handed neutrino $M_1<M_{\Delta_L}$, the 
$CP$-asymmetry parameter (\ref{maj-tri-asy}) can be rewritten as 
\be
\epsilon_1^{II}=\left( \frac{3M_1}{16\pi M_{\Delta_L}^2} \right) 
\frac{Im \left[ \left( (m_D^a)^\dagger M_R (m_D^b)^*\right)_{11}\beta_{ab} 
\right]} {\left( (m_D^a)^\dagger m_D^a\right)_{11}}\,. 
\label{final-typeII-asym}
\ee
We further assume $m_D\simeq \tilde{m}_D$ and $\beta=O(1)$. Thus 
using the value of $m_D$ and $M_R$ from equations (\ref{neu_dirac_mass}) 
and (\ref{maj-mass-matrix}) in the above equation we get
\be
\epsilon_1^{II}\simeq \left(-\frac{3M_1 \beta c^2 m_\mu^2}
{8\pi M_{\Delta_L}^2} \right) \frac{\epsilon \sin \delta}{2} 
\left(\frac{1}{m_1}-\frac{1}{m_2}\right)\,.
\label{cp-working-form}
\ee
\begin{figure}[htbp]
\begin{center}
\epsfig{file=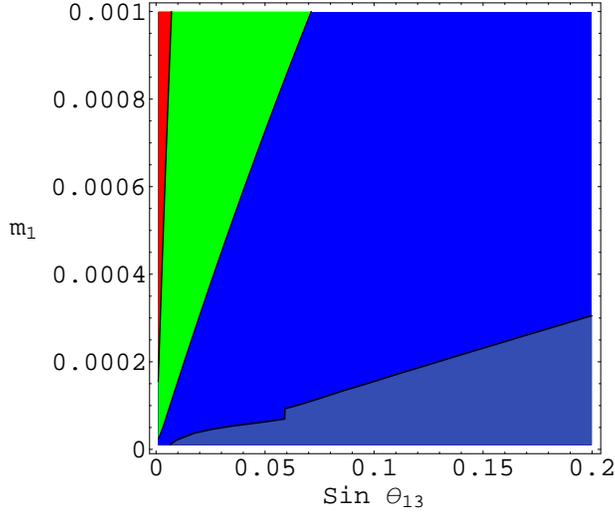}
\caption{Contours satisfying the required $B$-asymmetry in the 
$\sin \theta_{13}$ versus $m_1$ plane are plotted for 
$(4.2\times10^8 GeV/M_1)=0.01,0.001,0.0001$. We have used the 
parameters $\beta=1, c=1$ and $M_{\Delta_L}=10^{13}GeV$.}
\label{typeII-m1_theta13}
\end{center}
\end{figure}

Following the same procedure in section (\ref{dominate-typeI}) we 
calculate the $B$-asymmetry by using $\epsilon_1^{II}$. The corresponding 
regions in the $\sin \theta_{13}$ versus $m_1$ plane are shown in 
figure (\ref{typeII-m1_theta13}) for various values of $M_1$. In the 
bottom most region we have $4.2\times10^9 GeV$ $< M_1 <$ $4.2\times10^9$ 
GeV. As we go up the mass of $N_1$ increases by an order of magnitude 
for each region. By taking the best-fit value for $m_2=0.009$ eV and 
using the maximal $CP$-violation it is found that in a large allowed 
range of $\sin \theta_{13}$ the smaller values of $M_1$ are preferable 
for all $m_1<10^{-3}$ eV. That means a successful leptogenesis with 
$m_1<10^{-3}eV$ prefers the only values $4.2\times 10^{8} GeV\leq 
M_1< 4.2\times 10^{12} GeV$. Note that these regions are exactly 
complementary to the dominant type-I case. 

\subsection{$L$-asymmetry with $M_{\Delta_L}<M_1$} 
We now assume that $M_{\Delta_L}<M_1$. Hence the final 
$L$-asymmetry must be given by the decay of triplet $\Delta_L$.
The $L$-asymmetry from the decay of triplet $\Delta_L$ is defined as
\be
Y_L=\epsilon_\Delta Y_\Delta d\,,
\label{lepasy-def}
\ee
where $Y_\Delta=(n_{\Delta_L}/s)$, with $n_{\Delta_L}=n_{{\Delta_L}^{++}}+ 
n_{{\Delta_L}^{+}}+n_{{\Delta_L}^{0}}$ is the density of the triplets and 
$s$ is the entropy density, and $d$ is the dilution factor. Assuming 
$\beta$'s of order unity and substituting $\epsilon_\Delta$ from 
equation (\ref{triplet-asy}) we get the $L$-asymmetry 
\be
Y_L=\frac{1}{8\pi v^2} \frac{Im\left( Tr[(m_\nu^I)^*M_R]\sum\beta_i^*\right)}
{\sum |\beta_i|^2 r^2} Y_\Delta d\,.
\label{triplet-working-form}
\ee
\begin{figure}[htbp]
\begin{center}
\epsfig{file=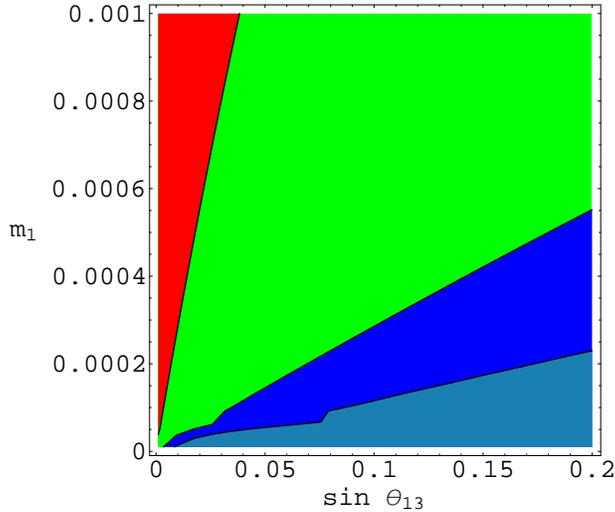}
\caption{Contours satisfying the required $B$-asymmetry in the 
$\sin \theta_{13}$ versus $m_1$ plane are shown for $r^2=1,10,25$. We 
have used the parameters $c=0.1$, $\beta=1$.}
\label{triplet-m1_theta13}
\end{center}
\end{figure}

Using the equations (\ref{neu-masses}) and  (\ref{maj-mass-matrix}) 
we evaluate $Y_L$. Again following the same procedure as given in 
section (\ref{dominate-typeI}) we calculate the $B$-asymmetry. With 
the maximal $CP$-violation and using the best-fit parameters, 
$m_2=0.009$ eV and $m_3=0.05$ eV, the regions in the $\sin \theta_{13}$ 
versus $m_1$ plane are shown in fig. (\ref{triplet-m1_theta13}) for 
various values of $r^2=v_R^2/M_{\Delta_L}^2$. In the bottom most region 
we have $r^2>25$. $r^2$ values are decreased further to wards upper-left 
(the red region which is not allowed because it represents $r^2<1$ which 
implies $M_{\Delta_L}>M_1$). Thus it is clear that for 
$\sin \theta_{13}< 0.2$ the only values of $m_1<10^{-4}$ eV are 
allowed for a successful leptogenesis.

\subsection{Results and Discussions}
Assuming the neutrino Dirac mass matrix follows the same hierarchy 
of charged lepton mass matrix we studied the sensitivity of 
$L$-asymmetry on the mass scale of the lightest right handed neutrino 
as well as the triplet $\Delta_L$. In any case it is found that 
a successful L-asymmetry requires the mass of lightest right handed 
neutrino should satisfy $M_1>O(10^8) GeV$ and that of $M_{\Delta_L}> 
O(10^{10})GeV$. Therefore, these mechanisms of producing L-asymmetry 
is far away from our hope to be verified in the next generation 
accelerators. On the other hand, the large masses of $N_1$ and $\Delta_L$ 
satisfy a large range of parameters explored in the neutrino oscillations. 
In the following we study an alternative to explain the 
$L$-asymmetry at the TeV scale that is compatible with the low 
energy neutrino oscillation data.     

\section{Transient left-right domain walls, leptogenesis and TeV scale 
right handed neutrino}
\subsection{Spontaneous breaking of D-parity and transient
left-right domain walls}
In the conventional low energy left-right symmetric model the 
discrete left-right symmetry as well as the guage symmetry 
$SU(2)_L\times SU(2)_R\times U(1)_{B-L}$ breaks at the same scale 
through the vev of $\Delta_R$. As a result stable domain 
walls~\cite{domain_wall}, interpolating between the  L and 
R-like regions, are formed. By L-like we mean regions favored by the 
observed phenomenology, while in the R-like regions the vacuum expectation 
value of $\Delta_R$ is zero. Unless some non-trivial mechanism 
prevents this domain structure, the existence of R-like domains 
would disagree with low energy phenomenology. Furthermore, the domain 
walls would quickly come to dominate the energy density of the 
Universe. Thus in this theory a departure from exact 
$L \leftrightarrow R$ symmetry is essential in such a way as to 
eliminate the phenomenologically disfavoured R-like regions. 

The domain walls formed can be {\it transient} if there exists a
slight deviation from exact $L \leftrightarrow R$ symmetry. In other 
words we require $g_L\neq g_R$ before $SU(2)_L\times SU(2)_R$ 
breaking scale. In the present case this is achieved by breaking the 
$D$-parity at a high scale, at around  $\eta_P\sim 10^{13}$ GeV. This 
gives rise to $g_L\neq g_R$ before the breaking of guage symmetry 
$SU(2)_L\times SU(2)_R$. As a result the spectrum of 
Higgs bosons exhibit the left-right {\it asymmetry} even though $SU(2)_R$ 
symmetry is unbroken. Therefore, the thermal perturbative corrections to 
the Higgs field free energy will not be symmetric and the domain walls 
will be unstable. The slight difference in the free energy between the 
two types of regions causes a pressure difference across the walls, 
converting all the R-like regions to L-like regions. Details of this 
dynamics can be found in ref.~\cite{cline&yajnik_prd.02}.

\subsection{Leptogenesis from transient domain walls}
It was shown in~\cite{cline&yajnik_prd.02} that within the thickness of 
the domain walls the net $CP$ violating phase becomes position dependent. 
Under these circumstances the preferred scattering of $\nu_L$ over its 
$CP$-conjugate state ($\nu_L^c$) produce a net raw 
$L$-asymmetry~\cite{cline&yajnik_prd.02}
\be
\eta^{\rm raw}_{\sss L} \cong 0.01\,  v_w {1\over g_*}\,
        {M_1^4\over T^5\Delta_w}
\label{eq:ans2}
\ee
where $\eta^{\rm raw}_{\sss L}$ is the ratio of $n_L$ to the entropy
density $s$. In the right hand side $\Delta_w$ is the wall width and
$g_*$ is the effective thermodynamic degrees of freedom at the epoch
with temperature $T$. Using $M_1 =f_1 \Delta_{\sss T}$, with 
$\Delta_{\sss T}$ is the temperature dependent $vev$ acquired by
the $\Delta_{\sss R}$ in the phase of interest, and $\Delta_w^{-1} 
=\sqrt{\lambda_{eff}}\Delta_{\sss T}$ in equation (\ref{eq:ans2}) we get
\be
\eta^{\rm raw}_L \cong 10^{-4} v_w
\left(\frac{\Delta_{\sss T}}{T} \right)^5 f_1^4
\sqrt{\lambda_{eff}}\,,
\ee
where we have used $g_*=110$. Therefore, depending on the various
dimensionless couplings, the raw asymmetry may lie in the range
$O(10^{-4}~-~10^{-10})$. However, it may not be the final $L$-asymmetry, 
because the thermally equilibrated $L$-violating processes mediated by 
the right handed neutrinos can erase the produced raw asymmetry. 
Therefore, a final $L$-asymmetry and hence the bound on right handed 
neutrino masses can only be obtained by solving the Boltzmann 
equations~\cite{baryo_lepto_group}. We assume a normal mass 
hierarchy in the right handed neutrino sector. In this scenario, as 
the temperature falls, first $N_3$ and $N_2$ go out of thermal 
equilibrium while $N_1$ is in thermal equilibrium. Therefore, it is 
the number density and mass of $N_1$ that are important in the present 
case which enter into the Boltzmann equations. The relevant Boltzmann 
equations for the present purpose are~\cite{sahu&yaj_prd.05,sahu&yaj_plb.06}
\bea
\frac{dY_{N1}}{dZ} &=& -(D+S)\left(Y_{N1}-Y^{eq}_{N1}\right)
\label{boltzmann.1}\\
\frac{dY_{B-L}}{dZ} &=& -W Y_{B-L}
\label{boltzmann.2},
\eea
where $Y_{N_1}$ is the density of $N_1$ in a comoving volume,
$Y_{B-L}$ is the $B-L$ asymmetry and the parameter $Z=M_1/T$. The
various terms $D$,$S$ and $W$ are representing the decay, scatterings
and the wash out processes involving the right handed neutrinos. In 
particular, $D=\Gamma_D/ZH$, with
\be
\Gamma_D=\frac{1}{16 \pi v^2}\tilde{m}_1 M_1^2,
\label{decay}
\ee
where $\tilde{m}_1=(m_D^{\dagger}m_D)_{11}/M_1$ is called the effective
neutrino mass parameter. Similarly $S=\Gamma_S/HZ$ and $W=\Gamma_W/HZ$. 
Here $\Gamma_S$ and $\Gamma_W$ receives the contribution from $\Delta_L=1$
and $\Delta_L=2$ L-violating scattering processes. 

In an expanding Universe these $\Gamma$'s compete with the Hubble 
expansion parameter. In a comoving volume the dependence of 
$\Delta_{\rm L}=1$ $L$-violating processes on the parameters 
$\tilde{m}_1$ and $M_1$ is given as 
\be
\left(\frac{\gamma_{D}}{sH(M_1)}\right), \left(\frac{
\gamma^{N1}_{\phi,s}}{sH(M_1)}\right), \left(\frac{
\gamma^{N1}_{\phi,t}}{sH(M_1)}\right) \propto k_1\tilde{m}_1\,.
\label{dilution}
\ee
On the other hand, the dependence of the $\gamma$'s in
$\Delta_{\rm L}=2$ L-number violating processes on $\tilde{m}_1$ 
and $M_1$ is given by
\be
\left(\frac{\gamma^l_{N1}}{sH(M_1)}\right), \left(\frac{
\gamma^l_{N1,t}}{sH(M_1)}\right) \propto k_2 \tilde{m}_1^2 M_1.
\label{washout}
\ee
Finally there are also $L$-conserving processes whose dependence 
is given by
\be
\left(\frac{\gamma_{Z'}}{sH(M_1)}\right) \propto k_3 M_1^{-1}\,.
\label{l-conserve}
\ee
In the above equations (\ref{dilution}), (\ref{washout}),
(\ref{l-conserve}), $k_i$, $i=1,2,3$ are dimensionful constants
determined from other parameters. Since the $L$-conserving
processes are inversely proportional to the mass scale of $N_1$,
they rapidly bring the species $N_1$ into thermal equilibrium
for all $T\gg M_1$. Furthermore, smaller the values of $M_1$, the
washout effects (\ref{washout}) are negligible because of their
linear dependence on $M_1$. We shall work in this regime
while solving the Boltzmann equations.
                                                                               
The equations (\ref{boltzmann.1}) and (\ref{boltzmann.2}) are
solved numerically. The initial $B-L$ asymmetry is the net raw
asymmetry produced through the domain wall mechanism as discussed 
above. We impose the following initial conditions:
\be
Y^{in}_{N1}=Y^{eq}_{N1}~~ {\mathrm and}~~ Y^{in}_{B-L}=\eta^{raw}_{B-L},
\label{initial-cond}
\ee
assuming that there are no other processes creating $L$-asymmetry 
below the $B-L$ symmetry breaking scale. This requires $\Gamma_D\leq 
H$ at an epoch $T\geq M_1$ and hence lead to a bound~\cite{fglp_bound}
\be
m_\nu < m_* \equiv 4\pi g_*^{1/2}\frac{G_N^{1/2}}{\sqrt{2}G_F}
= 6.5\times10^{-4}eV.
\label{eff_par_cons}
\ee
Alternatively in terms of Yukawa couplings this bound reads
\be
h_{\nu}\leq 10 x, ~~~{\rm with}~~ x=(M_1/M_{pl})^{1/2}.
\label{yukawa_const}
\ee
\begin{figure}[ht]
\begin{center}
\epsfig{file=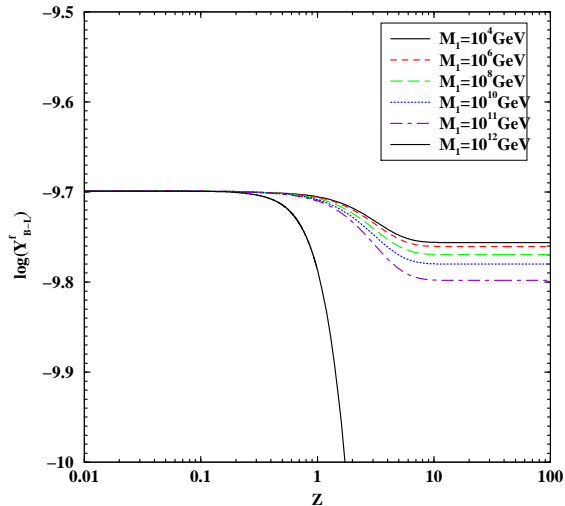, width=0.45\textwidth}
\caption{The evolution of $B-L$ asymmetry for different values of
$M_1$ shown against $Z(=M_1/T)$ for $\tilde{m}_1=10^{-4}$eV
and $\eta^{raw}_{B-L}=2.0\times 10^{-10}$}
\label{figure-1}
\end{center}
\end{figure}
At any temperature $T\geq M_1$, wash out processes involving $N_1$ are
kept under check due to the $\tilde{m}_1^2$ dependence in (\ref{washout})
for small values of $\tilde{m}_1$. As a result a given
raw asymmetry suffers limited erasure. As the temperature
falls below the mass scale of $N_1$ the wash out processes
become negligible leaving behind a final $L$-asymmetry.
Fig.\ref{figure-1} shows the result of solving the Boltzmann
equations for different values of $M_1$. An important conclusion
from this figure is that for smaller values of $M_1$ the wash out
effects are tiny. Hence by demanding that the initial raw asymmetry is
the required asymmetry of the present Universe we can conspire the
mass scale of $N_1$ to be as low as 1 TeV. For this value
of $M_1$, using equation (\ref{yukawa_const}), we get the constraint
on the neutrino Dirac Yukawa coupling to be $h_\nu\leq 10^{-7}$. It 
is shown in ref.~\cite{sahu&yaj_plb.06} that $h_\nu=10^{-7}$ is 
reasonable to suppress the flavor changing neutral current in 
the conventional left-right symmetric model.

We assume that in equation (\ref{boltzmann.2}) there are no other 
sources that produce $L$-asymmetry below the $SU(2)_R\times U(1)_{B-L}$ 
symmetry breaking phase transition. This can be justified by considering
small values of $h_\nu$, since the $CP$ asymmetry parameter $\epsilon_1$
depends quadratically on $h_\nu$. For $h_\nu\leq 10^{-7}$ the 
$L$-asymmetry $Y_L \leq O(10^{-14})$, which is far less than the raw 
asymmetry produced by the scatterings of neutrinos on the domain walls. 
This explains the absence of any $L$-asymmetry generating terms in 
equation (\ref{boltzmann.2}).

\section{Spontaneous breaking of D-parity and implications for 
cosmology}
An important aspect of the particle physics models is that the 
out-of-equilibrium decay of heavy scalar condensations gives rise 
to density perturbations in the early Universe~\cite{density_per}. 
In such a scenario, the cosmic microwave background radiation (CMBR) 
originating from the decay products of the scalar condensation and 
hence its anisotropy can be affected by the fluctuation of the 
scalar condensates. The observed anisotropy then constrain the mass 
scale of the heavy Higgs which induces the density perturbations. 
In the present model the fluctuation of the amplitude of late 
decaying condensation $\sigma$ (the so called curvaton scenario) 
can give rise to density perturbations if the energy density of $\sigma$ 
dominates the Universe for some time before its decay. Thus the models 
where inflaton doesn't generate sufficient perturbations can be 
rescued. 

One possibility is that the $\sigma$ can be abundantly produced 
from the decay of inflaton field and dominates before its decay. 
Note that $\sigma$ is a singlet field under the gauge symmetry 
$SU(2)_L\times SU(2)_R\times U(1)_{B-L}$. Therefore, the domination 
of $\sigma$ before it's decay is natural in this model than any 
other scalar fields which have the gauge interactions. This is 
possible if $\Gamma_{inf}\gg \Gamma_\sigma$, where $\Gamma_{inf}$ 
and $\Gamma_\sigma$ are respectively the decay rates of inflaton and 
$\sigma$ fields. The Universe will then go through a radiation
dominated era with a reheating temperature $T_I\simeq
g_*^{-1/4}(M_{pl}\Gamma_{inf})^{1/2}$ when the inflaton field
decays completely, i.e. $\Gamma_{inf}\sim H$. If the initial 
amplitude of $\sigma$ is substantial then it will reheat the 
Universe at a latter epoch $H \sim \Gamma_\sigma$ characterised 
by the reheat temperature $T_{II}\simeq g_*^{-1/4}(M_{pl}
\Gamma_{\sigma})^{1/2}$ when $\sigma$ decays completely. 
Therefore, the final density perturbation is mostly given by the 
$\sigma$ field~\cite{density_per}.    
  
Obtaining an acceptable perturbations of the correct size (about 1 in 
$10^5$) requires that the $vev$ of $\sigma$ field $\eta_P\sim 10^5 
H_I$~\cite{density_per}, where $H_I$ is the Hubble expansion rate 
during inflation. For $\eta_P\sim 10^{13}$ GeV (which is required to 
suppress the type-II contribution of the neutrino mass matrix) one can 
have $H_I\sim 10^8$ GeV. 

\section{Conclusions}
We studied BVL from the decay of right handed heavy Majorana 
neutrinos as well as the triplet $\Delta_L$ in a class of 
left-right symmetric models with spontaneous $D$-parity violation. 
While in a generic type-I seesaw models, assuming normal mass 
hierarchy in the right handed neutrino sector, one requires  
$M_1>4.2\times 10^8 GeV$ for successful thermal leptogenesis, with 
$D$-parity this bound can be lowered up to a factor of 
$\left(M_{\Delta_L}^2/4M\eta_P\right)$. Thus the lowering 
factor depends on the model parameters in the present case. On 
the other hand, in the case $M_{\Delta_L}<M_1$ the lower bound 
on the mass scale of $\Delta_L$ is of the order $10^{10}$ GeV 
to produce the required lepton asymmetry. In any case the 
thermal leptogenesis scale can not be lowered up to a TeV scale if 
the lepton asymmetry is produced through the out-of-equilibrium decay 
of these heavy particles (either right handed neutrinos or 
triplet Higgses). However, this is not true if the production and 
decay channel of these heavy particles in a thermal bath are 
different.   

The large masses of these heavy particles satisfy a large range of 
low energy neutrino oscillation data as we saw in figs. (
\ref{typeI-m1_theta13}), 
(\ref{typeII-m1_theta13}) and (\ref{triplet-m1_theta13}). In particular, 
we found that in case $M_1< M_{\Delta_L}$ (1) the dominating $\epsilon_1$ 
favors $M_1> 4.2\times 10^{12}$ GeV for all $m_1<10^{-3}$ eV, 
(2) the dominating $\epsilon_1^{II}$, on the other hand, favors 
$4.2\times 10^{8} GeV\leq M_1 < 4.2\times 10^{12} GeV$ for all 
$m_1<10^{-3}$ eV. In the case $M_{\Delta_L}<M_1$ we found that 
$m_1<10^{-4}$ eV are the only allowed values to give rise a 
successful leptogenesis.

Despite the success, the out-of-equilibrium decay production of 
$L$-asymmetry suffers a serious problem as far as the collider 
energy concern. Therefore, we considered an alternative 
mechanism of producing $L$-asymmetry by considering the extra source 
of $CP$-violation in the model. In particular, the complex condensate 
inside left-right domain wall gives rise to $CP$-violation. Under 
these circumstance the preferred scattering of $\nu_L$ over it's 
$CP$-conjugate state $\nu_L^c$ produce a net $L$-asymmetry. The survival of 
this asymmetry then requires the mass scale of $N_1$ to be very small, 
say $10 TeV$. This is compatible with the low energy neutrino oscillation 
data if the Dirac mass matrix of the neutrinos follow two orders of 
magnitude less than the charged lepton mass matrix. Moreover, the TeV 
scale masses of the right handed neutrinos are explained through the 
breaking of $SU(2)_R$ guage symmetry at a few TeV scale while leaving 
the $D$-parity breaking scale as high as $10^{13}$ GeV. 

Since $\sigma$ is a singlet scalar field under the gauge symmetry 
$SU(2)_L\times SU(2)_R\times U(1)_{B-L}$, we conjecture that its 
late decay can produce a density perturbation in the early 
Universe. However, in this work, we have not explored the details 
of density perturbations due to its out of equilibrium decay. This 
is under investigation and will be reported else where. 

\section*{Acknowledgment}
We thank Prof. Anjan Joshipura for useful discussions and suggestions.

\end{document}